\documentclass[11pt]{article}
\usepackage[symbol]{footmisc}
\def\correspondingauthor{\footnote{Corresponding author.  }}
\usepackage[left=2cm,top=2cm,right=2cm,bottom=2cm]{geometry}
\usepackage{amsmath, amssymb}

\usepackage{graphicx}
\usepackage{graphics}
\usepackage{epstopdf}
\usepackage{subfigure}
\usepackage{amsfonts}
\usepackage{sectsty}
\usepackage{sectsty}
\usepackage{hyperref,lineno}
\usepackage{cite}

\begin{document}
	%\setpagewiselinenumbers
%	\modulolinenumbers[1]
%	\linenumbers
	\begin{center}
		\large{\bf{Wormhole modeling in $R^2$ gravity with linear trace term}} \\
		\vspace{5mm}
		\normalsize{Nisha Godani$^1$ and Gauranga C. Samanta$^{2}{}$\correspondingauthor{} }\\
		\normalsize{$^1$Department of Mathematics, Institute of Applied Sciences and Humanities\\
		GLA University, Mathura, Uttar Pradesh, India\\
			$^2$Department of Mathematics, BITS Pilani K K Birla Goa Campus, Goa, India}\\
			\normalsize {nishagodani.dei@gmail.com\\gauranga81@gmail.com}
	\end{center}
\begin{abstract}
Morris and Thorne \cite{morris1} proposed traversable wormholes, hypothetical connecting tools,
using the concept of Einstein's general theory of relativity. In this paper, the modification of
general relativity (in particular $f(R,T)$ theory of gravity defined by Harko et al. \cite{harko})
is considered, to study the traversable wormhole solutions. The function $f(R,T)$ is considered as
$f(R,T)=R+\alpha R^2+\beta T$, where $\alpha$ and $\beta$ are controlling parameters.
The shape and red shift functions appearing in the metric of wormhole structure have significant contribution
in the development of wormhole solutions. We have considered both variable and constant red shift functions with
a logarithmic shape function. The energy conditions are examined, geometric configuration is analyzed and the
radius of the throat is determined in order to have wormhole solutions in absence of exotic matter.
\end{abstract}

\section{Introduction}
Wormholes are hypothetical geometrical structures which have a characteristic to connect two distinct space-times
or two distinct points of the same space-times. Flamm \cite{flamm} proposed the concept of wormhole. After Flamm,
a similar geometrical structure was obtained by  Einstein and Rosen \cite{eins-ros} which is known as Einstein-Rosen
bridge. The notion of traversable wormholes was first defined by Morris and Thorne \cite{morris1}  as a medium for
teaching general relativity. They obtained wormhole solutions in general relativity using a spherically symmetric
metric dependent of shape and red shift functions. Their solutions were filled with  the matter that does not obey
the energy conditions. Indeed, the traversable wormhole solutions may not be obtained in general relativity, if the
null energy condition is satisfied. This issue can be resolved by considering the systems where quantum effects
compete with the classical ones \cite{Visser1, Gao, Maldacena, Caceres, Fu}.
Further, quantum scalar stress energy tensor is used to obtain a self consistent solution of the semiclassical Einstein field equations corresponding to a Lorentzian wormhole \cite{Taylor}. Its back reaction problem is studied in \cite{Hochberg}. Specific solutions are found to represent a wormhole connecting two asymptotically flat regions \cite{Taylor} which indicates the possibility of inducing primordial wormholes at the early universe. Considering one loop effective action in large $N$ and $s$-wave approximations, an analytical solution is obtained for self consistent primordial wormhole with constant radius \cite{Nojiri99}. Further, for some initial conditions, GUTs at the early universe are used to obtain primordial wormholes at the early universe \cite{GUT}.
 Alternatively, the presence of additional
fields can also be considered as a source for the dissatisfaction of null energy condition (NEC) which has an association
with various problematic instabilities \cite{Bronnikov1, Armendariz-Picon, Dubovsky, Bronnikov2, Gonzalez, Gonzalez1,
Rubakov, Rubakov1, Evseev}.
The validation of energy conditions in the context of wormholes is a significant issue which has been dealt with
in literature, for instance in dynamic and thin shell wormholes \cite{Kar, Arellano, Cataldo, Mehdizadeh}, by
proposing new methods.
Other than this, various researchers have tried to obtain wormhole solutions using the
background  of modified theories of gravity. These solutions are developed in Kaluza-Klein gravity,  Born-Infeld
theory, Brans-Dicke theory, Einstein-Gauss-Bonnet theory, Einstein-Cartan theory, scalar tensor theory etc.
\cite{Agnese, Nandi, Dzhunushaliev, Bronnikov4, Richarte, Leon, Lobo1, Sushkov, Eiroa, Zangeneh, Bronnikov5,
Shaikh, Bronnikov6, Mehdizadeh1, Bronnikov3}. \\

\noindent
In modified theories of gravity, the stress energy tensor is replaced with effective stress energy tensor that
contains curvature terms of higher order. The generalized theories of gravity are used to sort out the problem of
exotic matter in wormholes, to construct viable cosmological models of our universe, to explain the singularities etc.
The $f(R)$ theory of gravity is one of the modified theories in which the geometrical part is modified by replacing
Ricci scalar curvature $R$ in Lagrangian gravitation action by a general function $f(R)$. The field equations obtained
with respect to this theory are highly complex and possesses a larger set of solutions than general relativity.
This theory is also simplified in \cite{Bertolami16} that provides a coupling between the matter and function $f(R)$
leading towards an extra force that may justify the current accelerating scenario of the universe
\cite{Nojiri15, Bertolami46}.   Starobinsky \cite{Starobinsky99} presented the first model of inflation. Subsequently, Nojiri and Odintsov\cite{Nojiri2003O}  presented a
modified gravity, in which the positive power of the curvature term supports an inflationary epoch, while the terms with negative powers of curvature serve as
effective dark energy, supports current cosmic acceleration.
Many other cosmological models are studied from different aspects in the context of $f(R)$ gravity
\cite{Capozziello91, Bombacigno21, Sbis, Chen, Elizalde26, Elizalde25, Astashenok1, Miranda, Nascimento,
Odintsov20, Odintsov1, Nojiri56, Parth}.
The $f(R)$ theory of gravity has been extensively used in the investigation of wormhole solutions.
The static wormholes using the non-commutative geometry are developed \cite{Rahaman, Jamil}.
The junction conditions in  $f(R)$ gravity are  applied to build pure double layer bubbles and thin
shell \cite{Eiroa1, Eiroa2, Eiroa3, Eiroa4}. The cosmological development of wormhole solutions is explored
in \cite{Bhattacharya}. Dynamical wormholes without need of exotic matter and asymptotically tending to FLRW
universe are obtained in \cite{Bahamonde}. Lorentzian wormhole solutions are analyzed with viable $f(R)$
model in \cite{Pavlovic}.
Traversable wormhole solutions are constructed in gravity and higher order curvature terms are found to be
responsible for the dissatisfaction of NEC \cite{Lobo12}. Taking constant shape and red shift functions, the
energy conditions for wormhole geometries are examined in \cite{saiedi}. However, with novel and variable shape
function and constant red shift function, these are examined in \cite{peter, godani, Samanta19}. Further,
the efforts are put to obtain the wormhole solutions with less amount of exotic matter using viable $f(R)$
gravity models \cite{Godani19, Samantaepjc}. Wormholes are also studied form different points of
view in \cite{Kar2, Wang, Sahdev, Roman, Poisson, Barcelo, Gonz, Visser2, Gonz1, Sushkov1, Lobo, Lobo2, Bohmer1,
Dotti, Forghani, Heydarzade, Moradpour,godani1}.\\

\noindent
The motivation of this work is to construct a wormhole solution in $R^2$ gravity with linear trace term. The main
purpose of considering such
type of modification of general relativity is to avoid the presence of exotic matter at the throat of the wormhole.
Subsequently, we try to study the
important role of redshift function in wormhole geometry. Therefore, two different types of redshift functions are
considered. Eventually, we estimate the
suitable range of controlling parameters $\alpha$ and $\beta$ the coefficients of $R^2$ and $T$, for the construction
of wormhole throat without support of exotic matter. The organisation of the paper is as follows: in section-2
explicit form of field equations and its analytic solutions are presented, in section-3 results are discussed,
finally concluding remarks are given in section-4.

\section{Field Equations \& Wormhole Geometry}
The static and spherically symmetric  metric defining the wormhole structure  is
\begin{equation}\label{metric}
ds^2=-e^{2\Phi(r)}dt^2+\frac{dr^2}{1-b(r)/r} + r^2(d\theta^2+\sin^2\theta d\phi^2).
\end{equation}
The  function $\Phi (r)$ determines the gravitational redshift, hence it is called redshift function.
The wormhole solutions must satisfy Einstein's field equations and must possess a throat that joins two regions of
universe which are asymptotically flat. For a traversable wormhole, event horizon should not be present and the
effect of tidal gravitational forces should be very small on a traveler.
The functions $\Phi(r)$ and $b(r)$ are the functions of  radial coordinate $r$, which is a non-decreasing function.
Its minimum value is $r_0>0$, radius of the throat, and maximum value is $+\infty$.  The function $b(r)$ determines
the shape of wormhole, hence it is called as shape function. The existence of wormhole solutions demands the satisfaction
of following conditions:
(i) $b(r_0)=r_0$, (ii) $\frac{b(r)-b'(r)r}{b^2}>0$, (iii) $b'(r_0)-1\leq 0$, (iv) $\frac{b(r)}{r}<1$ for $r>r_0$ and
(v) $\frac{b(r)}{r}\rightarrow 0$ as $r\rightarrow\infty$.
Traversable wormholes were first studied in the framework of general relativity \cite{morris1}.  The $f(R,T)$
theory of gravity is a generalization of  general relativity which were introduced by Harko et al. \cite{harko}.
It  replaces the  gravitational action $R$ with an arbitrary function $f(R,T)$ of $R$ and $T$, where $R$ is Ricci
scalar and $T$ is the trace of the energy-momentum tensor. The  gravitational action for $f(R,T)$ theory of gravity
is defined as modified Einstein's general relativity by replacing $R$ with an arbitrary function $f(R,T)$ of $R$ and
$T$, where $T$ is the trace of the energy-momentum tensor. The gravitational  action is defined as
\begin{equation}\label{action2}
S_G=\dfrac{1}{16\pi}\int[f(R,T)+L_m]\sqrt{-g}d^4x.
\end{equation}
Let $\square \equiv- \triangledown^{\mu}\triangledown_{\nu}$ and
$\theta_{\mu\nu} = -2T_{\mu\nu} + g^{\mu\nu}L_m - 2 g^{\gamma\sigma}\dfrac{\partial^2 L_m}{\partial
g^{\mu\nu}\partial g^{\gamma\sigma}}$.
Taking $L_m=-p$,

\begin{equation}\label{theta}
\theta_{\mu\nu} = -2T_{\mu\nu} - p g^{\mu\nu}.
\end{equation}

\noindent
Varying action (\ref{action2}) with respect to the metric, field equations are

\begin{equation}\label{frt}
f(R,T)R_{\mu\nu} -\frac{1}{2}f(R,T)g_{\mu\nu} + (g_{\mu\nu}
\square
-\triangledown_\mu\triangledown_\nu)f_R(R,T)=8\pi T_{\mu\nu} - f_T(R,T) \theta_{\mu\nu},
\end{equation}
where $f_R(R,T) \equiv \dfrac{\partial f(R, T)}{\partial R}$ and
$ f_T(R,T) \equiv \dfrac{\partial f(R,T)}{\partial T}.$

\noindent
For $f(R,T)=R + \alpha R^2 +\beta T$, where $\alpha$ and $\beta$ are arbitrary constants, the gravitational
field equations from Eq.\eqref{frt} are obtained as

\begin{equation}
\rho=\frac{1}{2(\beta+8 \pi)(2\beta+8 \pi)}\Bigg[(5\beta+16\pi)A_1+\beta(A_2+A_3)\Bigg]
\end{equation}

\begin{equation}
p_t=\frac{1}{4(\beta+8 \pi)}\Bigg[-2A_2-16\pi\rho-2\beta\rho+\frac{(3\beta +16\pi)}{(2\beta +8\pi)} (A_1+A_2+A_3)\Bigg]
\end{equation}

\begin{equation}
p_r=\frac{(A_1+A_2+A_3)}{(2\beta+8\pi)}-\rho-2p_t
\end{equation}
where
\begin{eqnarray}
A_1&=&\frac{1}{2 r^2}\Bigg[\left({\Phi^{'}(r) } (3 b(r)+{b^{'}(r)} r-4 r)-2 r {\Phi^{'}(r) }^2 (r-b(r))+2 r
{\Phi^{''}(r) } (r-b(r))\right) \nonumber\\
&\times&\left(\frac{1}{r^2}\Bigg[2 \alpha  \left(2 r (r-b(r)) \left({\Phi^{'}(r) }^2+{\Phi^{''}(r) }\right)+{\Phi^{'}(r) }
(4 r-3 b(r))-{b^{'}(r)} (r {\Phi^{'}(r) }+2)\right)\Bigg]+1\right)\Bigg]\nonumber\\
&+&\left(1-\frac{b(r)}{r}\right) \left(2 \alpha  {\Phi^{'}(r) }+\frac{\alpha  ({b^{'}(r)} r-b(r))}{r^2
\left(1-\frac{b(r)}{r}\right)}+\frac{4 \alpha }{r}\right)+\frac{1}{2} \left(-\frac{1}{r^4}\Bigg(\alpha
\left(2 r (r-b(r)) \right.\right.\nonumber\\
&\times&
\left.\left.\left({\Phi^{'}(r) }^2+{\Phi^{''}(r) }\right)+{\Phi^{'}(r) } (4 r-3 b(r))-{b^{'}(r)} (r {\Phi^{'}(r)
}+2)\right)^2\Bigg)-\frac{1}{r^2}\Bigg[2 r (r-b(r))
	\left({\Phi^{'}(r) }^2+{\Phi^{''}(r) }\right)\right.\nonumber\\
	&+&\left.{\Phi^{'}(r) } (4 r-3 b(r))-{b^{'}(r)} (r {\Phi^{'}(r) }+2)\Bigg]\right)-2 \alpha  {\Phi^{'}(r) }
	\left(1-\frac{b(r)}{r}\right)
\end{eqnarray}

\begin{eqnarray}
A_2&=&-\left(1-\frac{b(r)}{r}\right) \left(2 \alpha  {\Phi^{'}(r) }+\frac{\alpha  ({b^{'}(r)} r-b(r))}{r^2
\left(1-\frac{b(r)}{r}\right)}+\frac{4 \alpha }{r}\right)+\frac{\alpha  ({b^{'}(r)} r-b(r))}{r^2}\nonumber\\
&+&\frac{1}{2} \left(\frac{\alpha  \left(2 r (r-b(r)) \left({\Phi^{'}(r) }^2+{\Phi^{''}(r) }\right)+{\Phi^{'}(r) }
(4 r-3 b(r))-{b^{'}(r)} (r {\Phi^{'}(r) }+2)\right)^2}{r^4}\right.\nonumber\\
&+&\left.\frac{2 r (r-b(r)) \left({\Phi^{'}(r) }^2+{\Phi^{''}(r) }\right)+{\Phi^{'}(r) } (4 r-3 b(r))-{b^{'}(r)}
(r {\Phi^{'}(r) }+2)}{r^2}\right)\nonumber\\
&-&\frac{1}{2 r^3}\Bigg[\left(b(r) \left(-2 r^2 {\Phi^{'}(r) }^2-2 r^2 {\Phi^{''}(r) }+r {\Phi^{'}(r) }+2\right)
-{b^{'}(r)} r (r {\Phi^{'}(r) }+2)+2 r^3 \left({\Phi^{'}(r) }^2+{\Phi^{''}(r) }\right)\right)\nonumber\\
&\times&\left(\frac{2 \alpha  \left(2 r (r-b(r)) \left({\Phi^{'}(r) }^2+{\Phi^{''}(r) }\right)+{\Phi^{'}(r) }
(4 r-3 b(r))-{b^{'}(r)} (r {\Phi^{'}(r) }+2)\right)}{r^2}+1\right)\Bigg]
\end{eqnarray}

\begin{eqnarray}
A_3&=&-\frac{1}{r^2}\Bigg[\frac{1}{r^2}\left(-\frac{b(r) (2 r {\Phi^{'}(r) }+1)}{2 r}-\frac{{b^{'}(r)}}{2}+r
{\Phi^{'}(r) }\right) \left(\Bigg[2 \alpha  \left(2 r (r-b(r)) \left({\Phi^{'}(r) }^2+{\Phi^{''}(r) }\right)
\right.\right.\nonumber\\
&+&\left.\left.{\Phi^{'}(r) } (4 r-3 b(r))-{b^{'}(r)} (r {\Phi^{'}(r) }+2)\right)\Bigg]+1\right)\Bigg]-
\left(1-\frac{b(r)}{r}\right) \left(2 \alpha  {\Phi^{'}(r) }+\frac{\alpha  ({b^{'}(r)} r-b(r))}{r^2
\left(1-\frac{b(r)}{r}\right)}\right.\nonumber\\
&+&\left.\frac{4 \alpha }{r}\right)+\frac{1}{2} \left(\frac{\alpha  \left(2 r (r-b(r)) \left({\Phi^{'}(r)
}^2+{\Phi^{''}(r) }\right)+{\Phi^{'}(r) } (4 r-3 b(r))-{b^{'}(r)} (r {\Phi^{'}(r) }+2)\right)^2}{r^4}\right.\nonumber\\
&+&\left.\frac{2 r (r-b(r)) \left({\Phi^{'}(r) }^2+{\Phi^{''}(r) }\right)+{\Phi^{'}(r) } (4 r-3 b(r))-{b^{'}(r)}
(r {\Phi 1}+2)}{r^2}\right)+\frac{2 \alpha  \left(1-\frac{b(r)}{r}\right)}{r}
\end{eqnarray}

\noindent
In the present study, we considered two red shift functions (i) $\Phi(r)=c$ (constant) and (ii)
$\Phi(r)=\frac{1}{r}$ with shape function $b(r)=\frac{r_0\log(r+1)}{\log(r_0+1)}$ and obtained the expressions
for energy density and energy condition terms which are as follows:
\\

\noindent
\textbf{Case 1:} $\Phi(r)=c$ (constant)\\
\begin{eqnarray}
\rho&=&\frac{1}{8 (\beta +4 \pi ) (\beta +8 \pi ) r^5 (r+1)^2 \log ^2(r_0+1)}\Bigg[4 \alpha  (7 \beta +32 \pi )
(r+1)^2 r^4 \log ^2(r_0+1)\nonumber\\
&+&(r+1) r^3 r_0 \log (r_0+1) (\beta  (8 \alpha  r+9)+32 \pi  (\alpha  r+1))-(r+1) r_0 \log (r+1) \left(r^2 (r+1)
\right.\nonumber\\
&\times& \left.\log (r_0+1) (\beta +36 \alpha  \beta  r+160 \pi  \alpha  r)-4 \alpha  \beta  r_0\right)-8 \alpha
(3 \beta +8 \pi ) r r_0^2\Bigg]
\end{eqnarray}

\begin{eqnarray}
\rho+p_r&=&-\frac{r_0 ((r+1) \log (r+1)-r) \left(r^2 (r+1) (\alpha  r+1) \log (r_0+1)-4 \alpha  r_0\right)}{(\beta +8
\pi ) r^5 (r+1)^2 \log ^2(r_0+1)}
\end{eqnarray}

\begin{eqnarray}
\rho+p_t&=&\frac{1}{4 (\beta +8 \pi ) r^5 (r+1)^2 \log ^2(r_0+1)}\Bigg[12 \alpha  (r+1)^2 r^4 \log ^2(r_0+1)+(r+1)
r^3 r_0\nonumber\\
&\times&  (2 \alpha  r+3) \log (r_0+1)-(r+1) r_0 \log (r+1) \left((r+1) r^2 (14 \alpha  r-1) \log (r_0+1)\right.\nonumber\\
&+&\left.4 \alpha  r_0\right)-8 \alpha  r r_0^2\Bigg]
\end{eqnarray}

\begin{eqnarray}
\rho+p_r+2p_t&=&\frac{1}{4 (\beta +4 \pi ) r^5 (r+1)^2 \log ^2(r_0+1)}\Bigg[-4 \alpha  (r+1)^2 r^4 \log ^2(r_0+1)\nonumber\\
&+&(r+1) r^3 r_0 \log (r_0+1)+(r+1) r_0 \log (r+1) \left((r+1) r^2 (4 \alpha  r-1)\right. \nonumber\\
&\times&\left.\log (r_0+1)+4 \alpha  r_0\right)-8 \alpha  r r_0^2\Bigg]
\end{eqnarray}

\begin{eqnarray}
\rho-|p_r|&=&\frac{1}{8 (\beta +4 \pi ) (\beta +8 \pi ) r^5 (r+1)^2 \log ^2(r_0+1)}\Bigg[4 \alpha  (7 \beta +32 \pi )
(r+1)^2 r^4 \log ^2(r_0+1)\nonumber\\
&+&(r+1) r^3 r_0 \log (r_0+1) (\beta  (8 \alpha  r+9)+32 \pi  (\alpha  r+1))-(r+1) r_0 \log (r+1) \left(r^2 (r+1)\right.
\nonumber\\
&\times& \left.\log (r_0+1) (\beta +36 \alpha  \beta  r+160 \pi  \alpha  r)-4 \alpha  \beta  r_0\right)-8 \alpha
(3 \beta +8 \pi ) r r_0^2\Bigg]\nonumber\\
&-&\Bigg| \frac{1}{8 (\beta +4 \pi ) (\beta +8 \pi ) r^5 (r+1)^2 \log ^2(r_0+1)}\Bigg[(7 \beta +32 \pi ) (r+1)
r_0 \log (r+1) \left((r+1) r^2 \right.\nonumber\\
&\times& \left.(4 \alpha  r-1) \log (r_0+1)+4 \alpha  r_0\right)-r \left(4 \alpha  (7 \beta +32 \pi ) (r+1)^2
r^3 \log ^2(r_0+1)\right.\nonumber\\
&+& \left.\beta  (r+1) r^2 r_0 \log (r_0+1)+8 \alpha  (\beta +8 \pi ) r_0^2\right)\Bigg] \Bigg|
\end{eqnarray}
\begin{eqnarray}
\rho-|p_t|&=&\frac{1}{8 (\beta +4 \pi ) (\beta +8 \pi ) r^5 (r+1)^2 \log ^2(r_0+1)}\Bigg[4 \alpha  (7 \beta +32 \pi )
(r+1)^2 r^4 \log ^2(r_0+1)\nonumber\\
&+&(r+1) r^3 r_0 \log (r_0+1) (\beta  (8 \alpha  r+9)+32 \pi  (\alpha  r+1))-(r+1) r_0 \log (r+1) \left(r^2 (r+1)\right.
\nonumber\\
&\times& \left.\log (r_0+1) (\beta +36 \alpha  \beta  r+160 \pi  \alpha  r)-4 \alpha  \beta  r_0\right)-8 \alpha
(3 \beta +8 \pi ) r r_0^2\Bigg]\nonumber\\
&-&\Bigg| \frac{1}{8 (\beta +4 \pi ) (\beta +8 \pi ) r^5 (r+1)^2 \log ^2(r_0+1)}\Bigg[-4 \alpha  (\beta +8 \pi )
(r+1)^2 r^4 \log ^2(r_0+1)\nonumber\\
&-&(r+1) r^3 r_0 \log (r_0+1) (\beta  (4 \alpha  r+3)+8 \pi  (2 \alpha  r+1))+(r+1) r_0 \log (r+1) \left(r^2 (r+1)
\right.\nonumber\\
&\times& \left. \log (r_0+1) (\beta  (8 \alpha  r+3)+8 \pi  (6 \alpha  r+1))-4 \alpha  (3 \beta +8 \pi ) r_0\right)+
8 \alpha  \beta  r r_0^2\Bigg] \Bigg|
\end{eqnarray}

\textbf{Case 2:} $\Phi(r)=\frac{1}{r}$
\begin{eqnarray}
\rho&=&\frac{1}{8 (\beta +4 \pi ) (\beta +8 \pi ) r^{10} (r+1)^2 \log ^2(r_0+1)}\Bigg[\alpha  (r+1)^2 (r+2) r_0^2
\left(\beta  \left(2 r^2+61 r-30\right)\right.\nonumber\\
&+&\left.16 \pi  (13 r-6)\right) \log ^2(r+1)+r^2 \left(\alpha  \left(-r^2\right) (2 r-1) r_0^2 (3 \beta
(4 r-5)+16 \pi  (2 r-3))\right.\nonumber\\
&+&\left.2 (r+1)^2 \log ^2(r_0+1) \left(32 \pi  \left(-3 \alpha +2 \alpha  r^7+2 r^5-r^4+8 \alpha  r\right)+\beta
\left(-30 \alpha +14 \alpha  r^7\right.\right.\right.\nonumber\\
&+&\left.\left.\left.4 \alpha  r^6+19 r^5-9 r^4+76 \alpha  r\right)\right)+r (r+1) r_0 \log (r_0+1) \left(\beta
\left(-60 \alpha +8 \alpha  r^6+9 r^5-9 r^4\right.\right.\right.\nonumber\\
&-&\left.\left.\left.152 \alpha  r^2+160 \alpha  r\right)+32 \pi  \left(-6 \alpha +\alpha  r^6+r^5-r^4-16 \alpha
r^2+16 \alpha  r\right)\right)\right)-(r+1) r r_0\nonumber\\
 &\times&\log (r+1) \left((r+1) \log (r_0+1) \left(32 \pi  \left(-12 \alpha +5 \alpha  r^7+3 r^5-2 r^4+8 \alpha
 r^2+26 \alpha  r\right)\right.\right.\nonumber\\
 &+&\left.\left.\beta  \left(-120 \alpha +36 \alpha  r^7+(8 \alpha +1) r^6+29 r^5-18 r^4+80 \alpha  r^2+244 \alpha
 r\right)\right)-2 \alpha  r r_0 \left(16 \pi \right.\right. \nonumber\\
 &\times&\left.\left.
 \left(12 r^2-13 r+6\right)+\beta  \left(2 r^3+54 r^2-65 r+30\right)\right)\right)\Bigg]
\end{eqnarray}

\begin{eqnarray}
\rho+p_r&=&\frac{1}{(\beta +8 \pi ) r^{10} (r+1)^2 \log ^2(r_0+1)}\Bigg[\left(r (r r_0+2 (r+1) \log (r_0+1))
\right.-\left(r^2+3 r+2\right)\nonumber\\
&\times&\left. r_0 \log (r+1)\right) \left(2 \alpha  \left(-r^3-2 r^2+r+2\right) r_0 \log (r+1)+r \left((r+1)
\left(-4 \alpha +\alpha  r^6\right.\right.\right.\nonumber\\
&+&\left.\left.\left.r^5-r^4+4 \alpha  r\right) \log (r_0+1)-2 \alpha  r \left(2 r^2-3 r+1\right) r_0\right)\right)\Bigg]
\end{eqnarray}

\begin{eqnarray}
\rho+p_t&=&-\frac{1}{4 (\beta +8 \pi ) r^{10} (r+1)^2 \log ^2(r_0+1)}\Bigg[\alpha  (r+1)^2 \left(2 r^3-27 r^2-52
r+20\right) r_0^2 \nonumber\\
&\times&\log ^2(r+1)+r^2 \left(-\left(\alpha  r^2 \left(-8 r^2+14 r-5\right) r_0^2+2 (r+1)^2 \left(-10 \alpha
+6 \alpha  r^7+2 \alpha  r^6\right.\right.\right.\nonumber\\
&+&\left.\left.\left.9 r^5-3 r^4+36 \alpha  r\right) \log ^2(r_0+1)+r (r+1) r_0 \left(-20 \alpha +2 \alpha
r^6+3 r^5-3 r^4-72 \alpha  r^2\right.\right.\right.\nonumber\\
&+&\left.\left.\left.64 \alpha  r\right) \log (r_0+1)\right)\right)+(r+1) r r_0 \log (r+1) \left(2 \alpha
r \left(2 r^3-30 r^2+27 r-10\right) r_0(r+1)\right.\nonumber\\
&\times&\left. \left(-40 \alpha +14 \alpha  r^7+(4 \alpha -1) r^6+15 r^5-6 r^4+32 \alpha  r^2+124 \alpha
r\right) \log (r_0+1)\right)\Bigg]
\end{eqnarray}

\begin{eqnarray}
\rho+p_r+2p_t&=&\frac{1}{4 (\beta +4 \pi ) r^{10} (r+1)^2 \log ^2(r_0+1)}\Bigg[\alpha  (r+1)^2 \left(2 r^3+13
r^2+12 r-12\right) r_0^2\nonumber\\
&\times& \log ^2(r+1)+r^2 \left(\alpha  r^2 \left(-8 r^2+10 r-3\right) r_0^2+r (r+1) r_0 \left(-12 \alpha
+r^5-r^4\right.\right.\nonumber\\
&-&\left.\left.24 \alpha  r^2+32 \alpha  r\right) \log (r_0+1)-2 (r+1)^2 \left(6 \alpha +2 \alpha  r^7-4 \alpha
r^6-3 r^5+r^4-12 \alpha  r\right)\right.\nonumber\\
&\times&\left. \log ^2(r_0+1)\right)+(r+1) r r_0 \log (r+1) \left(2 \alpha  r \left(2 r^3+6 r^2-13 r+6\right)
r_0+(r+1) \right.\nonumber\\
&\times&\left.\left(24 \alpha +4 \alpha  r^7-(8 \alpha +1) r^6-5 r^5+2 r^4-16 \alpha  r^2-36 \alpha  r\right)
\log (r_0+1)\right)\Bigg]
\end{eqnarray}

\begin{eqnarray}
\rho-|p_r|&=&\frac{1}{8 (\beta +4 \pi ) (\beta +8 \pi ) r^{10} (r+1)^2 \log ^2(r_0+1)}\Bigg[\alpha
(r+1)^2 (r+2) r_0^2 \left(\beta  \left(2 r^2+61 r-30\right)\right.\nonumber\\
&+&\left.16 \pi  (13 r-6)\right) \log ^2(r+1)+r^2 \left(\alpha  \left(-r^2\right) (2 r-1) r_0^2
(3 \beta  (4 r-5)+16 \pi  (2 r-3))\right.\nonumber\\
&+&\left.2 (r+1)^2 \log ^2(r_0+1) \left(32 \pi  \left(-3 \alpha +2 \alpha  r^7+2 r^5-r^4+8 \alpha
r\right)+\beta  \left(-30 \alpha +14 \alpha  r^7\right.\right.\right.\nonumber\\
&+&\left.\left.\left.4 \alpha  r^6+19 r^5-9 r^4+76 \alpha  r\right)\right)+r (r+1) r_0 \log (r_0+1)
\left(\beta  \left(-60 \alpha +8 \alpha  r^6+9 r^5-9 r^4\right.\right.\right.\nonumber\\
&-&\left.\left.\left.152 \alpha  r^2+160 \alpha  r\right)+32 \pi  \left(-6 \alpha +\alpha  r^6+r^5-r^4-16
\alpha  r^2+16 \alpha  r\right)\right)\right)-(r+1) r r_0\nonumber\\
&\times&\log (r+1) \left((r+1) \log (r_0+1) \left(32 \pi  \left(-12 \alpha +5 \alpha  r^7+3 r^5-2 r^4+8
\alpha  r^2+26 \alpha  r\right)\right.\right.\nonumber\\
&+&\left.\left.\beta  \left(-120 \alpha +36 \alpha  r^7+(8 \alpha +1) r^6+29 r^5-18 r^4+80 \alpha  r^2+244
\alpha  r\right)\right)-2 \alpha  r r_0 \left(16 \pi \right.\right. \nonumber\\
&\times&\left.\left.
\left(12 r^2-13 r+6\right)+\beta  \left(2 r^3+54 r^2-65 r+30\right)\right)\right)\Bigg]\nonumber\\
&-&\Bigg| \frac{1}{8 (\beta +4 \pi ) (\beta +8 \pi ) r^{10} (r+1)^2 \log ^2(r_0+1)}\Bigg[\alpha  (r+1)^2 (r+2)
r_0^2 \left(\beta  \left(14 r^2-45 r-2\right)\right.\nonumber\\
&+&\left.16 \pi  \left(4 r^2-9 r-2\right)\right) \log ^2(r+1)+r^2 \left(-\left(\alpha  r^2 (2 r-1) r_0^2 (\beta
(4 r-1)+16 \pi  (2 r-1))\right.\right.\nonumber\\
&+&\left.\left.r (r+1) r_0 \log (r_0+1) \left(\beta  \left(4 \alpha +r^5-r^4-88 \alpha  r^2+32 \alpha  r\right)
-64 \pi  \alpha  \left(4 r^2-1\right)\right)\right.\right.\nonumber\\
&+&\left.\left.2 (r+1)^2 \log ^2(r_0+1) \left(32 \pi  \left(\alpha +2 \alpha  r^7-\alpha  r^6+r^5+4 \alpha
r\right)+\beta  \left(2 \alpha +14 \alpha  r^7-4 \alpha  r^6\right.\right.\right.\right.\nonumber\\
&+&\left.\left.\left.\left.+11 r^5-r^4+44 \alpha  r\right)\right)\right)\right)+(r+1) r r_0 \log (r+1)
\left(2 \alpha  r r_0 \left(\beta  \left(14 r^3-54 r^2+17 r+2\right)\right.\right.\nonumber\\
&+&\left.\left.16 \pi  \left(4 r^3-12 r^2+r+2\right)\right)+(r+1) \log (r_0+1) \left(32 \pi  \left(4
\alpha +4 \alpha  r^7-(2 \alpha +1) r^6\right.\right.\right.\nonumber\\
&+&\left.\left.\left.2 r^5+18 \alpha  r\right)+\beta  \left(8 \alpha +28 \alpha  r^7-(8 \alpha +7)
r^6+21 r^5-2 r^4+16 \alpha  r^2+180 \alpha  r\right)\right)\right)\Bigg]\Bigg|
\end{eqnarray}

\begin{eqnarray}
\rho-|p_t|&=&\frac{1}{8 (\beta +4 \pi ) (\beta +8 \pi ) r^{10} (r+1)^2 \log ^2(r_0+1)}\Bigg[\alpha  (r+1)^2
(r+2) r_0^2 \left(\beta  \left(2 r^2+61 r-30\right)\right.\nonumber\\
&+&\left.16 \pi  (13 r-6)\right) \log ^2(r+1)+r^2 \left(\alpha  \left(-r^2\right) (2 r-1) r_0^2 (3 \beta
(4 r-5)+16 \pi  (2 r-3))\right.\nonumber\\
&+&\left.2 (r+1)^2 \log ^2(r_0+1) \left(32 \pi  \left(-3 \alpha +2 \alpha  r^7+2 r^5-r^4+8 \alpha  r\right)+
\beta  \left(-30 \alpha +14 \alpha  r^7\right.\right.\right.\nonumber\\
&+&\left.\left.\left.4 \alpha  r^6+19 r^5-9 r^4+76 \alpha  r\right)\right)+r (r+1) r_0 \log (r_0+1) \left(\beta
\left(-60 \alpha +8 \alpha  r^6+9 r^5-9 r^4\right.\right.\right.\nonumber\\
&-&\left.\left.\left.152 \alpha  r^2+160 \alpha  r\right)+32 \pi  \left(-6 \alpha +\alpha  r^6+r^5-r^4-16 \alpha
r^2+16 \alpha  r\right)\right)\right)-(r+1) r r_0\nonumber\\
&\times&\log (r+1) \left((r+1) \log (r_0+1) \left(32 \pi  \left(-12 \alpha +5 \alpha  r^7+3 r^5-2 r^4+8 \alpha
r^2+26 \alpha  r\right)\right.\right.\nonumber\\
&+&\left.\left.\beta  \left(-120 \alpha +36 \alpha  r^7+(8 \alpha +1) r^6+29 r^5-18 r^4+80 \alpha  r^2+244 \alpha
r\right)\right)-2 \alpha  r r_0 \left(16 \pi \right.\right. \nonumber\\
&\times&\left.\left.
\left(12 r^2-13 r+6\right)+\beta  \left(2 r^3+54 r^2-65 r+30\right)\right)\right)\Bigg]\nonumber\\
&-&\Bigg|   \frac{1}{8 (\beta +4 \pi ) (\beta +8 \pi ) r^{10} (r+1)^2 \log ^2(r_0+1)}\Bigg[-\alpha  (r+1)^2 (r+2)
r_0^2 \left(\beta  \left(6 r^2-r-10\right)\right.\nonumber\\
&+&\left.8 \pi  \left(2 r^2-5 r-2\right)\right) \log ^2(r+1)+r^2 \left(\alpha  r^2 (2 r-1) r_0^2 (\beta  (4 r-5)-
8 \pi )-2 (r+1)^2 \right.\nonumber\\
&\times&\left.\log ^2(r_0+1) \left(\beta  \left(-10 \alpha +2 \alpha  r^7+r^5-3 r^4+4 \alpha  r\right)+8 \pi
\left(-2 \alpha +2 \alpha  r^7-2 \alpha  r^6-r^5
-r^4\right.\right.\right.\nonumber\\
&-&\left.\left.\left.4 \alpha  r\right)\right)-r (r+1) r_0 \log (r_0+1) \left(\beta  \left(-20 \alpha +4 \alpha
r^6+3 r^5-3 r^4-8 \alpha  r^2+32 \alpha  r\right)\right.\right.\nonumber\\
&+&\left.\left.8 \pi  \left(-4 \alpha +2 \alpha  r^6+r^5-r^4+8 \alpha  r^2\right)\right)\right)+(r+1) r r_0 \log
(r+1) \left((r+1) \log (r_0+1) \right. \nonumber\\
&\times&\left.\left(8 \pi  \left(-8 \alpha +6 \alpha  r^7+(1-4 \alpha ) r^6-3 r^5-2 r^4-20 \alpha  r\right)+\beta
\left(-40 \alpha +8 \alpha  r^7+3 r^6-r^5-6 r^4\right.\right.\right.\nonumber\\
&+&\left.\left.\left.16 \alpha  r^2-4 \alpha  r\right)\right)-2 \alpha  r r_0 \left(\beta  \left(6 r^3-6 r^2-11
r+10\right)+8 \pi  \left(2 r^3-6 r^2+r+2\right)\right)\right)\Bigg]\Bigg|
\end{eqnarray}

\section{Results}
The present article is focused on the study of traversable wormholes which were proposed by Morris and Thorne
\cite{morris1} to teach general theory of relativity. They obtained wormhole solutions in general relativity
(GR) which demand the existence of exotic matter, the matter not obeying the energy conditions. This outcome
has opened an area of research for the exploration of wormhole solutions without need of exotic matter. These
are extensively studied in generalized theories of gravity developed in literature. The $f(R,T)$ theory of gravity,
introduced by Harko et al. \cite{harko}, is one of these theories in which traversable wormholes are studied. In this
study, we have considered the model $f(R,T)=R+\alpha R^2 +\beta T$, where $R$ is Ricci scalar curvature, $T$ is the
trace of stress energy tensor, $\alpha$ and $\beta$ are arbitrary constants. The metric of wormhole is  dependent on
shape and red shift functions which is useful in describing its characteristics and obtaining wormhole solutions.
We assumed the shape function $b(r)=\frac{r_0\log(r+1)}{\log(r_0+1)}$ \cite{godani}, where $r_0$ denotes the radius
of throat of wormhole. Further, to avoid the existence of horizons, the red shift function  $\Phi(r)$ should be non-zero.
It can be constant or variable.  For simplicity, many authors have considered $\Phi(r)=c$ (constant). To achieve
asymptotically flat wormholes, It should satisfy the condition  $e^{\Phi(r)}\rightarrow 1$ as $r\rightarrow \infty$.
The choice  $\Phi(r)=\frac{1}{r}$ satisfies this condition. We have taken both $\Phi(r)=c$ and   $\Phi(r)=\frac{1}{r}$
and explored the regions where energy conditions (ECs) namely, null energy condition (NEC), weak energy condition
(WEC), strong energy condition (SEC) and dominant energy condition (DEC) are consistent. In terms of radial pressure
$p_r$ and tangential pressure $p_t$, these
ECs are defined in the following manner: (i) NEC is said to be satisfied if $\rho + p_r\geq 0$ and $\rho + p_t\geq 0$;
(ii) WEC is said to be obeyed if $\rho\geq 0$, $\rho + p_r\geq 0$ and $\rho + p_t\geq 0$; (iii) SEC is said to be
validated if $\rho + p_r\geq 0$, $\rho + p_t\geq 0$ and  $\rho + p_r +2p_t\geq 0$; (iv) DEC is said to be fulfilled
if $\rho\geq 0$, $\rho - \lvert p_r\rvert \geq 0$ and $\rho - \lvert p_t\rvert \geq 0$. Further, the equation of
state in terms of radial pressure is $p_r=\omega \rho$, where $\omega$ is called the equation of state parameter,
and the anisotropy parameter $\triangle$ in terms of pressures $p_t$ and $p_r$ is defined as $\triangle=p_t-p_r$
which can have positive, negative or zero value. The positive value of $\triangle$ indicates the repulsive nature of
the geometry, negative value suggests the attractive nature of the geometry and zero value tells that the geometry is
isotropic.
We have carried our study mainly in two cases:
Case 1: $\Phi(r)=c$ (constant) and Case 2: $\Phi(r)=\frac{1}{r}$. Each case is detected in the following four subcases:
(i)  $\alpha=0$, $\beta=0$; (ii) $\alpha=0$, $\beta\neq 0$;  (iii) $\alpha\neq 0$, $\beta=0$; (iv) $\alpha\neq 0$,
$\beta\neq 0$. The results are as follows:\\

\noindent
\textbf{Case 1:} $\Phi(r)=c$ (constant)\\
\noindent
\textbf{Subcase 1(i):} $\alpha=0$, $\beta=0$, i.e. $f(R,T)=R$\\
After putting $\alpha=\beta =0$, the modified gravity converted into general relativity. In this subcase,
first the nature of energy density $\rho$ with respect to radial coordinate $r$ is examined. It is found to
have positive values for every $r>0$. But the first NEC term takes negative values for every $r>0$, so NEC is
dissatisfied and hence the wormholes are completely filled with exotic matter. This case study indicates that the
presence of exotic
matter may not be possible to avoid under the specific choice of shape and redshift functions.\\

\noindent
\textbf{Subcase 1(ii):} $\alpha=0$, $\beta\neq0$, i.e. $f(R,T)=R+\beta T$\\
In this case, the model converted into linear scalar curvature term coupled with linear trace term. This subcase
depends on parameter $\beta$ which can be positive or negative. For $\beta>0$, the results are similar to Subcase 1(i)
and for $\beta<0$, $\rho<0$ for $r>0$. This subcase is not able to avoid the exotic matter for any range of $r$. So, it is of no interest.
Hence, the modification of general relativity with linear scalar curvature coupled with linear trace may not be
enough to avoid the existence of exotic matter at wormhole throat with constant redshift function.\\

\noindent
\textbf{Subcase 1(iii):} $\alpha\neq 0$, $\beta=0$, i.e. $f(R,T)=R+\alpha R^2$\\
This subcase depends on parameter $\alpha$ which can be positive or negative. For $\alpha>0$, $\rho>0$ for $r>0.9$;
first NEC term $\rho+p_r>0$ for $r<1.3$ and second NEC term $\rho+p_t>0$ for $r>1$. This means NEC as well as
WEC are satisfied for $r\in(1,1.3)$. The SEC term $\rho+p_r+2p_t$ is found to be negative for all $r>0$;
first DEC term $\rho-|p_r|>0$ for $r\in(1,1.3)$ and second DEC term $\rho-|p_t|>0$ for $r>1$. Thus, NEC, WEC
and DEC are satisfied for $r\in(1,1.3)$. Consequently, this subcase provides the existence of non-exotic matter
for a small range of $r$ near the throat.  In this range, we have $\omega<-1$ and $\triangle>0$, i.e. wormholes
are filled with phantom fluid and have repulsive geometry for $r\in(1,1.3)$. Further, if $\alpha<0$,
then $\rho>0$ for $0<r<1$ and  $\rho+p_r>0$ for $r>1.3$. So, none of the above energy condition is satisfied.
 \\

\noindent
\textbf{Subcase 1(iv):} $\alpha\neq0$,  $\beta\neq 0$, i.e. $f(R,T)=R+\alpha R^2 +\beta T$\\
 In this subcase, both parameters  $\alpha$ and $\beta$ are present which can be positive or negative.
 If $\alpha>0$, then, for $\beta<-25$, $\rho>0$, for $r<1$ and $\rho+p_r>0$ for $r>1.1$ which implies the
 violation of all energy conditions. For $\beta\geq -25$, $\rho>0$ for $r>0.8$, $\rho+p_r>0$ for $r<1.2$,
 $\rho+p_t>0$ for $r>1.2$. This means that no EC is valid for $\alpha>0$.  Further, let $\alpha<0$.
 Then for $\beta\geq-25$, $\rho>0$ for $r<1$ and $\rho+p_r>0$ for $r>1.4$ which implies the violation of all
 energy conditions. For $\beta< -25$, $\rho>0$ for $r>1$, $\rho+p_r>0$ for $r<1.4$, $\rho+p_t>0$ for $r>1$,
 $\rho-|p_r|>0$ for $r\in(1,1.4)$, $\rho-|p_t|>0$ for $r>1$ and $\rho+p_r+2p_t<0$ for $r>0$. This means that
 NEC, WEC and DEC are valid only for $r\in(1,1.4)$ otherwise all ECs are dissatisfied. For $r\in(1,1.4)$,
 $\omega<-1$ and $\triangle>0$ which shows that the geometric configuration is repulsive and filled with phantom fluid.
 Thus, we have obtained the validation of energy conditions for $r\in(1,1.2)$ with $\alpha>0$, $\beta\geq -25$ and
 for $r\in(1,1.4)$ with $\alpha<0$, $\beta<-25$.\\

 \noindent
 \textbf{Case 2:} $\Phi(r)=\frac{1}{r}$ \\
 \noindent
 \textbf{Subcase 2(i):} $\alpha=0$, $\beta=0$, i.e. $f(R,T)=R$\\
 In this subcase, $\rho>0$ for all $r>0$. The first NEC term $\rho+p_r>0$, for $r\geq 1$ and the second NEC
 term $\rho+p_t>0$, for all $r>0$. This means both NEC and WEC are obeyed for $r\geq 1$. Hence, this study indicates
 that, there could be possible to avoid the presence of exotic matter at the throat
 of the wormhole in general relativity by constructing suitable choice of variable redshift and shape functions.
 Therefore, redshift and shape functions may be played as an important role to avoid the presence of exotic matter
 at wormhole throat. Further, $\rho+p_r+2p_t<0$ for all $r>0$, $\rho-|p_r|>0$ for $r\geq 1$ and $\rho-|p_t|>0$
for $r>0.4$. Thus all NEC, WEC and DEC are obeyed for $r\geq 1$ and SEC is obeyed nowhere. The anisotropy parameter
is positive throughout i.e. the geometry is repulsive throughout. The equation of state parameter $\omega$ is negative
throughout. For $1\leq r<8.9$, $-1<\omega<0$ and for $r\geq 8.9$, $\omega\leq -1$.  This depicts the existence of
wormhole solutions filled with non-phantom fluid near the throat and with phantom fluid outside of the throat.\\

\noindent
 \textbf{Subcase 2(ii):} $\alpha=0$, $\beta\neq0$, i.e. $f(R,T)=R+\beta T$\\
 This subcase depends on the parameter $\beta$ which can be positive or negative. First, let $\beta>0$. Then we have
 found $\rho>0$ for all $r>0$. The first NEC term $\rho+p_r>0$, for $r\geq 0.85$ and the second NEC term $\rho+p_t>0$,
 for all $r>0$. This shows the validation of both NEC and WEC  for $r\geq 0.85$. Further,
 $\rho+p_r+2p_t>0$ for all $r>0$, $\rho-|p_r|>0$ for $r\geq 0.$ and $\rho-|p_t|>0$ for all $r>0$.
 Thus all NEC, WEC, SEC and DEC are obeyed for $r\geq 0.85$. Therefore, this case study indicates
 that the modification of general relativity could be more reliable to avoid the
 presence of exotic matter at the throat of the wormhole by considering the appropriate choice of redshift
 and shape functions. Subsequently, the anisotropy parameter is positive throughout i.e. the geometry is
 repulsive throughout. The equation of state parameter $\omega$ is negative throughout.
 For $0.85\leq r<8.9$, $-1<\omega<0$ and for $r\geq 8.9$, $\omega\leq -1$.  This depicts the existence
 of wormhole solutions filled with non-phantom fluid near the throat and with phantom fluid outside of the throat.
 Furthermore, if $\beta<0$, then we have found  $\rho<0$ for all $r>0$, which indicates that the existence of
 exotic matter throughout the wormhole geometry. Therefore, to avoid abnormal matter, in this particular model,
 $\beta>0$ must be required.  \\

\noindent
\textbf{Subcase 2(iii):} $\alpha\neq0$, $\beta=0$, i.e. $f(R,T)=R+\alpha R^2$\\
In this subcase, the model reduces to $f(R)$ gravity \cite{Starobinsky99}. The parameter $\alpha$ can take both
positive or negative values. First we assume $\alpha>0$. Then we have found $\rho>0$ for $r\geq 0.9$. Then the
first NEC term $\rho+p_r>0$, for $r\geq 0.2$ and the second NEC term $\rho+p_t>0$, for  $r>0.9$. This gives the
validity of both NEC and WEC  for $r\geq 0.9$. Further, $\rho+p_r+2p_t<0$ for all $r>0$, $\rho-|p_r|>0$ for
$r\geq 0.9$ and $\rho-|p_t|>0$ for all $r>0.9$. Thus all NEC, WEC and DEC are satisfied for $r\geq 0.9$,
however, the SEC is dissatisfied everywhere. The anisotropy parameter is positive for $r>0.9$ and negative
otherwise. Further, the equation of state parameter $\omega$ is negative throughout. For $0.9\leq r<8.9$,
$-1<\omega<0$ and for $r\geq 8.9$, $\omega\leq -1$.  Similar to Subcase 2(ii), wormholes are found to filled
with non-phantom fluid near the throat and with phantom fluid away from the throat. Further, let $\alpha<0$.
Then we have found  $\rho<0$ for all $r>0$. Hence, only $\alpha>0$ gives the desired results.
 Therefore, $\alpha>0$ could be a better choice in the Starobinsky\cite{Starobinsky99} model to avoid
 the presence of exotic matter at wormhole throat. \\

\noindent
\textbf{Subcase 2(iv):} $\alpha\neq0$,  $\beta\neq 0$, i.e. $f(R,T)=R+\alpha R^2 +\beta T$\\
 This subcase depends on both $\alpha$ and $\beta$ which can have positive or negative values.
 Let $\alpha$ be positive. Then for $\beta<-25$, $\rho>0$ for $r\leq 1$, the first NEC term $\rho+p_r>0$,
 for $r< 0.3$ and the second NEC term $\rho+p_t>0$, for  $r<1.2$. This gives the validity of both NEC and WEC
 for $r<0.3$. Further,    $\rho+p_r+2p_t>0$ for all $r<1.3$, $\rho-|p_r|<0$ for all $r>0$ and $\rho-|p_t|>0$
 for all $r\leq 1$. Thus all NEC, WEC and SEC are satisfied for $r < 0.3$ and DEC is violated everywhere. For
 $r<0.3$, the anisotropy parameter is positive and the equation of state parameter $\omega<-1$. Thus, we have
 got the validation of NEC, WEC and SEC  near the throat only. If $\beta>-25$, then $\rho>0$ for $r>0.7$. Then
 the first NEC term $\rho+p_r>0$, for $r> 0.2$ and the second NEC term $\rho+p_t>0$, for  $r>0.7$. This gives
 the validity of both NEC and WEC  for $r>0.7$. Further,    $\rho+p_r+2p_t<0$ for all $r>0$. Both $\rho-|p_r|$
 and $\rho-|p_t|$ are positive for all $r>0.7$. This implies the validation of NEC, WEC and DEC for $r>0.7$ and
 SEC is dissatisfied everywhere. The anisotropy parameter is positive for $r>0.7$ and negative otherwise.
 Further, the equation of state parameter $\omega$ is negative throughout. For $0.7\leq r<8.9$, $-1<\omega<0$
 and for $r\geq 8.9$, $\omega\leq -1$.
 Further, let $\alpha$ be negative. Then for $\beta<-25$, we have obtained the same results as for $\alpha>0$,
 $\beta\geq -25$.  For $\beta\geq-25$, the results are same as for $\alpha>0$, $\beta\geq -25$. Thus,
 this subcase provides the favorable results for $r>0.7$ with (a) $\alpha>0$, $\beta\geq -25$ and (b)
 $\alpha<0$, $\beta<-25$.\\

\noindent
Thus, in case of constant  redshift function, we have found the satisfaction of energy conditions near the
throat for (a) $r\in(1,1.3)$ with $\alpha>0$, $\beta=0$ and (b) $r\in(1,1.4)$ with $\alpha<0$, $\beta<-25$.
However, in case of variable red shift function, the validation of energy conditions is obtained for (a) $r\geq 1$
with $\alpha=0$, $\beta=0$; (b) $r \geq 0.85$ with $\alpha=0$, $\beta>0$; (c)  $r\geq 0.9$ with $\alpha>0$, $\beta=0$;
(d) $r>0.7$ with $\alpha>0$, $\beta\geq -25$ and (e) $r>0.7$ with $\alpha<0$, $\beta\geq -25$.  It can be observed
that ECs NEC, WEC and DEC are obeyed only for a small range of $r$ near the throat, if $\Phi(r)$ is taken to be constant.
However, these ECs are valid for large ranges of $r$ with $\Phi(r)=\frac{1}{r}$. The minimum and maximum values of
radius of throat are obtained as 0.7 and 1 respectively. Consequently for any real value of parameters $\alpha$ and
$\beta$, we can have wormhole solutions completely free from exotic matter, if the radius of throat is taken as unity or greater than of that.
Thus, we have found the ranges of parameters $\alpha$ and $\beta$ that provides the wormhole solutions with non-exotic matter.   \\
\begin{figure}
	\centering
	\subfigure[This figure indicates $\rho>0$ and close to zero for $r>0.7$]{\includegraphics[scale=.78]{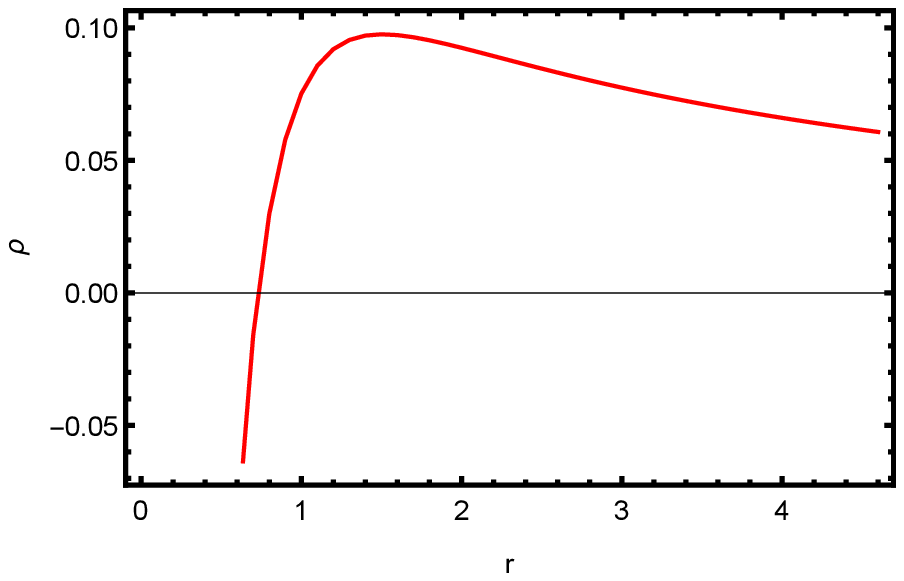}}\hspace{.1cm}
	\subfigure[This figure represents $\rho+p_r>0$ for $r> 0.2$ and, decreases rapidly as $r\to\infty$]{\includegraphics[scale=.78]{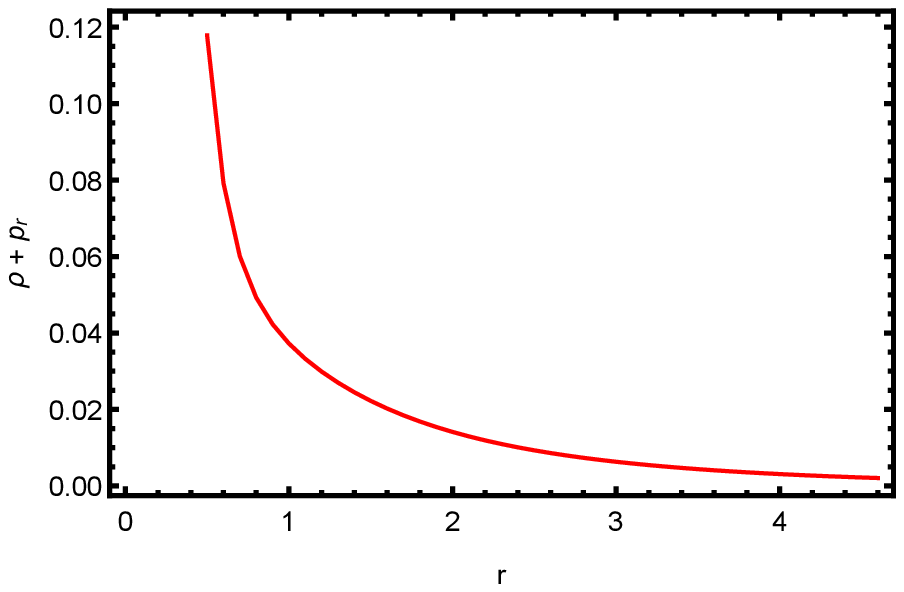}}\hspace{.1cm}
	\subfigure[From this figure it is observed that $\rho+p_t<0$ for $r<0.7$, however $\rho+p_t>0$ and decreases slowly for $r>0.7$ ]{\includegraphics[scale=.78]{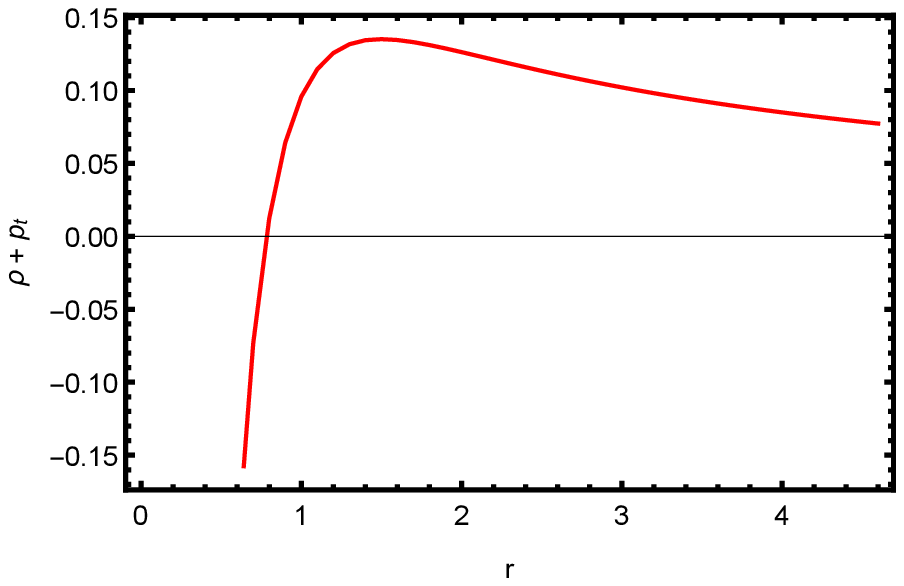}}\hspace{.1cm}
	\subfigure[$\rho-|p_r|>0$ for $r>0.7$ and decreases sharply for $r\to\infty$]{\includegraphics[scale=.78]{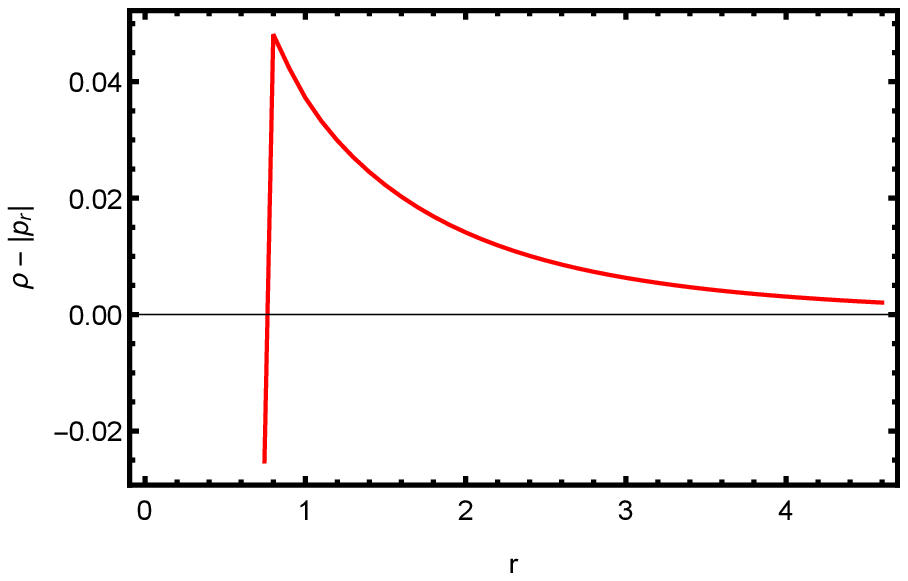}}\hspace{.1cm}
	\subfigure[$\rho-|p_t|>0$ for $r>0.7$ and decreases smoothly for $r\to\infty$ ]{\includegraphics[scale=.78]{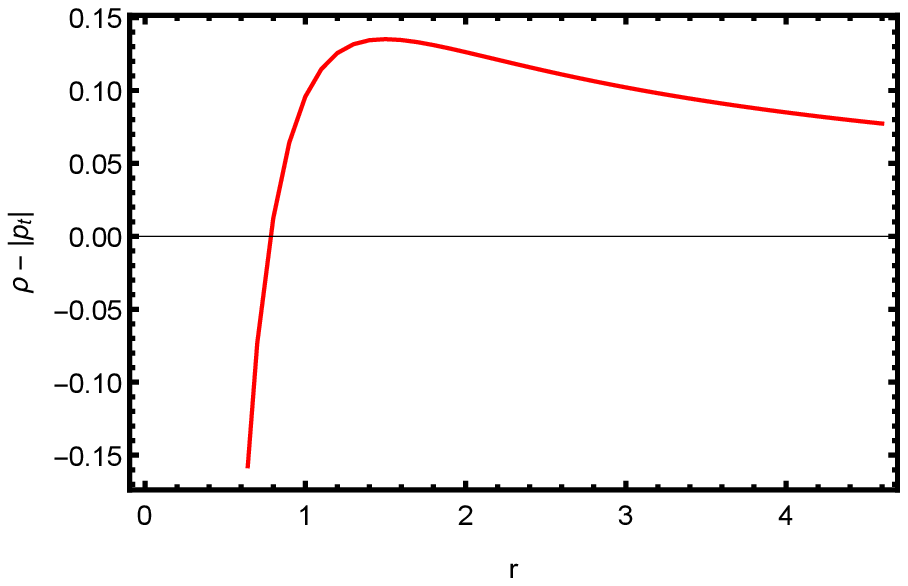}}\hspace{.1cm}
	\subfigure[This figure indicates the violation SEC, i. e. $\rho+p_r+2p_t<0$ for $r>0$]{\includegraphics[scale=.78]{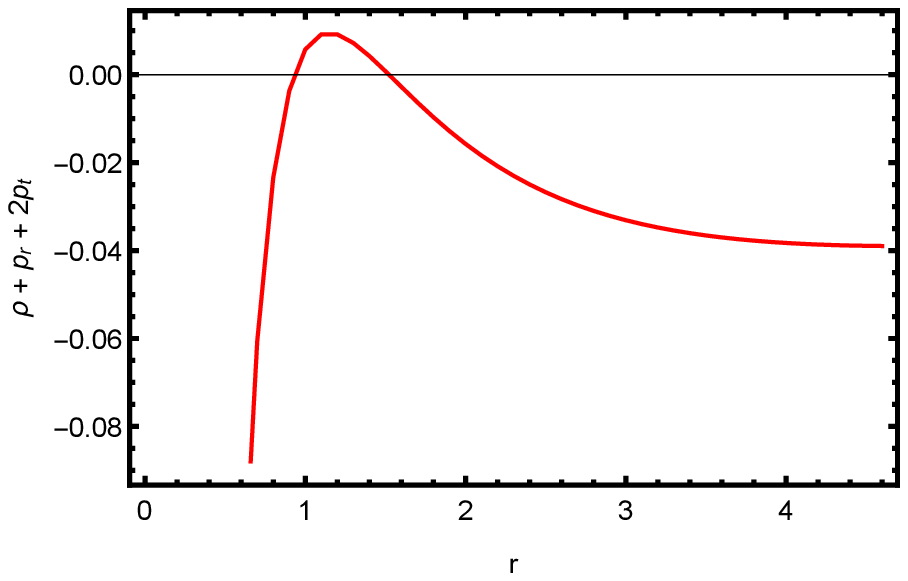}}\hspace{.1cm}
	\subfigure[The equation of state parameter $\omega$ is negative throughout.]{\includegraphics[scale=.78]{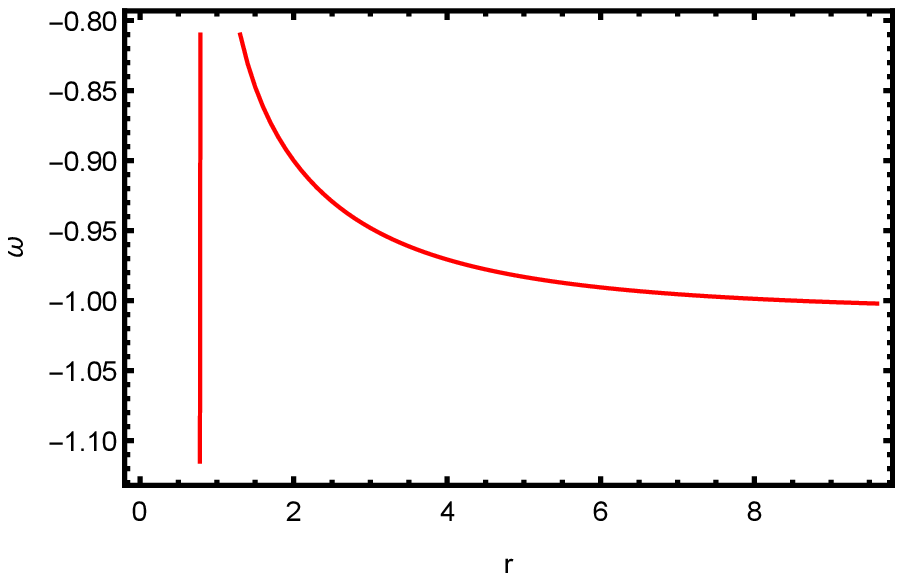}}\hspace{.1cm}
	\subfigure[The anisotropy parameter $\triangle>0$ for $r>0.7$ and negative otherwise. ]{\includegraphics[scale=.78]{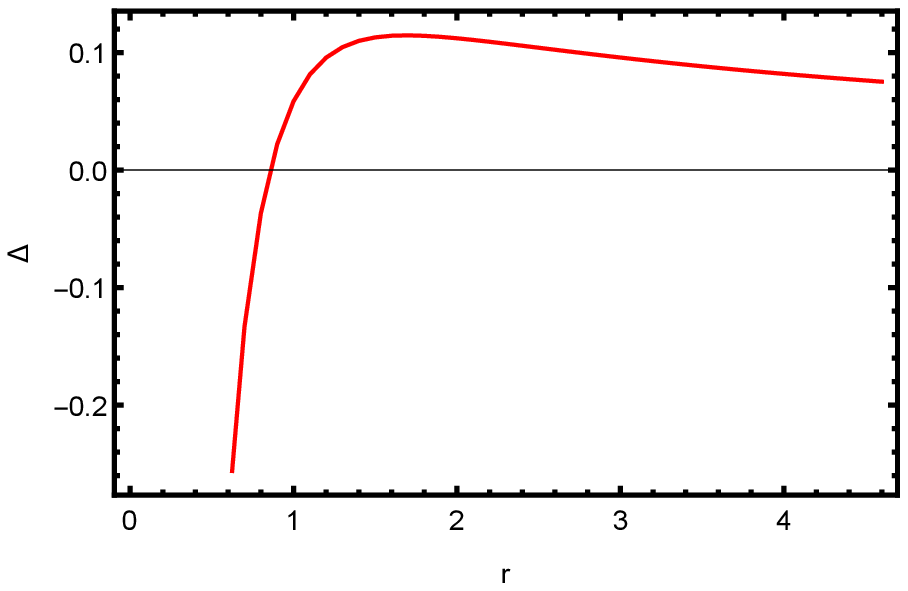}}\hspace{.1cm}
	%	\subfigure[$\rho+p_r$]{\includegraphics[scale=.45]{2n11}}\hspace{.1cm}
	%	\subfigure[$\rho-|p_r|$]{\includegraphics[scale=.45]{2d11}}\hspace{.1cm}
	\caption{Plots for Density, NEC, SEC, DEC, $\triangle$ \& $\omega$ with $\phi(r)=\frac{1}{r}$, $\alpha>0$ and $\beta>0$}
\end{figure}

\pagebreak
 %\noindent
 \section{Conclusion}
 In the present paper, $f(R,T)$ theory of gravity, a generalized theory, describing a coupling between matter and
 geometry is taken into account with the Lagrangian combination of quadratic in $R$ and linear in $T$.
 Precisely, $f(R,T)=R+\alpha R^2 +\beta T$ is considered, where $\alpha$ and $\beta$ are arbitrary constants,
 to explore traversable wormholes  introduced by Morris and Thorne \cite{morris1}. The aim of this work is to
 obtain wormhole solutions and determine the radius of the throat so that the presence of exotic matter could be ignored
 completely. Subsequently, we try to investigate the significant role of the redshift function in terms of constant and variable, shape function and modified gravity for the existence of wormhole solutions by avoiding the presence of exotic matter in the universe. To achieve this aim, first the wormhole solutions are obtained for shape function
 $b(r)=\frac{r\log(r_0+1)}{\log(r+1)}$ with both constant and variable red shift functions $\Phi(r)$.
 Then the validity of energy conditions is examined for each possible value of parameters $\alpha$ and $\beta$ in
 both cases. In case of constant $\Phi(r)$, the energy conditions are found to satisfy near the throat for a small
 range of $r$. These are obtained to be satisfied for  $r\in(1,1.3)$ with $\alpha>0$, $\beta=0$ and for $r\in(1,1.4)$
 with $\alpha<0$, $\beta<-25$. Therefore, the constant redshift function probably not a suitable choice to avoid the exotic matter for the wormhole solutions. Further, in case of variable redshift function, the validation of energy conditions is
 obtained for (a) $r\geq 1$ with $\alpha=0$, $\beta=0$; (b) $r \geq 0.85$ with $\alpha=0$, $\beta>0$; (c)  $r\geq 0.9$
 with $\alpha>0$, $\beta=0$; (d) $r>0.7$ with $\alpha>0$, $\beta\geq -25$ and (e) $r>0.7$ with $\alpha<0$, $\beta\geq -25$.
 Among all, the subcase-2(i) i. e. $r\geq 1$ with $\alpha=0$, $\beta=0$ is more interesting. If we substitute $\alpha=\beta=0$ in $f(R,T)=R+\alpha R^2 +\beta T$, then the model converted to general Einstein gravity. Morris and Thorne \cite{morris1} constructed traversable wormhole with constant redshift function and showed that exotic matter is one of the necessary component near the throat of the traversable wormhole. However, in this study, we may conclude, there could be possible to avoid the presence of exotic matter at the throat
 of the wormhole in general relativity by constructing suitable choice of variable redshift and shape functions. Consequently, it is observed that ECs are valid for large ranges of $r$  in second (variable redshift function) case. This shows a significant difference in the results
 for two distinct choices.  In spite of this, the geometric structure is analyzed to be repulsive in nature which is
 required near the throat.   For the model taken, if one considers the radius of throat greater than or equal to one,
 then the wormhole solutions without exotic matter could be obtained. We have found the existence of  wormholes filled with the matter satisfying the energy conditions.
 Our physical universe is also filled with the matter obeying the energy conditions. Consequently, this theoretically study supports and provides a tool to connect two distant objects of our physical universe.  Hence, the model undertaken favors the existence of
 wormhole solutions without any requirement of exotic matter with a suitable choice of redshift and shape in appropriate modified gravity.\\

 \noindent
 {\bf Acknowledgement:} The authors are very much thankful to the anonymous reviewer and editor for their constructive comments to improve the quality of work.

 %%%%%%%%%%%%%%%%%%%%%%%%%
%
 
\end{document}